\documentclass[hyper]{JHEP3} 

\usepackage{epsfig}

\newcommand\fverb{\setbox\pippobox=\hbox\bgroup\verb}

\newcommand\fverbdo{\egroup\medskip\noindent%

            \fbox{\unhbox\pippobox}\ }

\newcommand\fverbit{\egroup\item[\fbox{\unhbox\pippobox}]}

\newbox\pippobox

\title{BRST Invariance of 
Non-local Charges and Monodromy
Matrix of Bosonic String on 
$AdS_5\times S^5$}

\author{J. Kluso\v{n}
 \footnote{On leave from Masaryk University, Brno}\\
Dipartimento di Fisica \& Sezione I.N.F.N.\\
Universit\`a di Roma
``Tor Vergata'' \\
Via della Ricerca Scientifica 1 00133  Roma   ITALY\\
E-mail:
\email{Josef.Kluson@roma2.infn.it}}
\preprint{
\hepth{0701013}}
\abstract{Using the generalized Hamiltonian
method of Batalin, Fradkin and Vilkovsky we develop
the BRST formalism for the bosonic string
on $AdS_5\times S^5$ formulated as principal
chiral model. Then we
 show that the monodromy matrix
and non-local charges are BRST invariant.}


\newcommand{\mT}{\mathcal{T}}
\def\tr{\mathrm{Tr}}

\def\pb  #1{\left\{#1\right\}}

\newcommand{\mH}{\mathcal{H}}

\newcommand{\mD}{\mathcal{D}}

\newcommand{\mL}{\mathcal{L}}

\newcommand{\hJ}{\hat{J}}

\newcommand{\mY}{\mathcal{Y}}
\newcommand{\mZ}{\mathcal{Z}}

\newcommand{\mC}{\mathcal{C}}
\newcommand{\omC}{\overline{\mC}}
\newcommand{\mP}{\mathcal{P}}
\newcommand{\omP}{\overline{\mP}}

\begin{document}
\section{Introduction and Summary}
One of the most remarkable developments of
the string theory in recent years is the
celebrated duality between string theory on 
an $AdS_5\times S^5$ background and $N=4$
supersymmetric Yang-Mills theory living on the
boundary of this space
\cite{Maldacena:1997re,Witten:1998qj,Gubser:1998bc}
\footnote{For review,see for example
\cite{Aharony:1999ti,D'Hoker:2002aw}.}. These theories
have many properties in common, as 
for example global symmetries.
On the other hand the direct comparison of general
states in these two theories is difficult 
by the weak/strong nature of the duality and for
a long time could only be applied to the special
states that are protected by supersymmetry.

One of the most remarkable tools for checking this
duality is the discussion of integrability on 
both sides of this duality. This program began when
Minahan and Zarembo demonstrated that the one
loop anomalous dimension operator, acting on 
single trace scalar operators, could be interpreted
as the Hamiltonian of an integrable spin chain
\cite{Minahan:2002ve}  
\footnote{For review and extensive list of
references, see 
\cite{Minahan:2006sk,Plefka:2005bk,
Beisert:2004ry,Tseytlin:2003ii}.}.
Therefore, the anomalous dimensions could
be found using  Bethe ansatz. At present no
gauge theory calculation of these single trace
operators has contracted integrability.  

In string theory the investigation of the
integrability began with the discovery of
complete set of classically conserved non-local
charges in \cite{Bena:2003wd}
\footnote{For some of the related works, see
\cite{Mann:2006rh,Arutyunov:2006ak,
Dorey:2006mx,Gromov:2006dh,Frolov:2006cc,Dorey:2006zj,Alday:2005ww,
Das:2005hp,Arutyunov:2005nk,
Chen:2005uj,Alday:2005gi,Beisert:2005bm,Das:2004hy,Swanson:2004qa,
Arutyunov:2004yx,
Hatsuda:2004it,Kazakov:2004qf,Alday:2003zb}.}.
In summary, the evidence of the integrability 
on both sides of the duality is a compelling new
argument in favor of the duality.

The next step would be to extend the context
of the integrability on the quantum level. It turns
out that this is much more difficult problem,
for careful discussion, see \cite{Mann:2005ab}.
In the quantum treatment one should certainly
consider the Green-Schwarz superstring on 
the $AdS_5\times S^5$. On the other hand there
is no covariant quantization scheme for Green-Schwarz
superstring. Even if the light-cone treatment
of the Green-Schwarz string is very useful and
gives many interesting results 
 \cite{Metsaev:2000yu,Metsaev:2000yf} it would
 be certainly desirable to have covariant
 description of the superstring. Such a 
 formulation has been developed in last
 few years by Berkovits
\cite{Berkovits:2000fe,
Berkovits:2000ph,Berkovits:2000nn, Berkovits:2001us}
 \footnote{For
review of pure spinor formalism in 
superstring theory, see
\cite{Berkovits:2002zk,Grassi:2005av,
Grassi:2003cm,Grassi:2002sr,
Nekrasov:2005wg}.}. Pure spinor string 
in $AdS_5\times S^5$ has been
studied in 
 \cite{Berkovits:2004xu,Vallilo:2002mh,Berkovits:2000yr,
 Kluson:2006wq,Bianchi:2006im,Puletti:2006vb,Grassi:2006tj}
and the aspects of the classical
integrability have been discussed in 
\cite{Vallilo:2003nx,Berkovits:2004jw}.
On the other hand the explicit solution of the
pure spinor string in $AdS_5\times S^5$ background
is still lacking.

Since the BRST treatment of the string theory and
its relation to the integrability is not well
known we mean that it would be useful to study 
its properties in simpler dynamical system. 
One example of such a system is a bosonic string
theory on $AdS_5\times S^5$ formulated as
principal sigma model coupled to two dimensional 
world-sheet gravity. Due to the  diffeomorphism
invariance  this system possesses explicit
gauge invariance and can be treated with 
the Hamiltonian method of Batalin, Fradkin 
and Vilkovsky (BFV) \cite{Batalin:1977pb}.
We will develop this formalisms for the
principal sigma model coupled to two dimensional
gravity. We start with the Hamiltonian formalism
for corresponding principal model. We introduce
the canonical variables using the method
developed in  
\cite{Faddeev:1987ph}
\footnote{For recent application of this method
in string theory and quantum gravity,
see \cite{Das:2004hy,Das:2005hp,
Bianchi:2006im,Korotkin:1997fi,Miller:2006bu,
Kluson:2006ij}.}.
Then we introduce the extended phase space following
\cite{Fujiwara:1992eb}. We determine the 
action of the BRST charge
on the phase space variables using the
canonical Poisson brackets. We obtain, by definition,
expressions that hold for fields that are off-shell.
On the other hand it is well known that the
monodromy matrix is defined using the 
Lax connection that is constructed from currents
that define given principal model. In other words
the monodromy matrix is defined in the configuration
space. For that reason we use the field 
redefinitions  given in \cite{Fujiwara:1992eb}
that map the ghosts given in 
BFV formalism to ones that appear in the geometrical
treatment. It turns out however that even in this
case the BRST transformations of the currents 
do not correspond to the geometrical one.
We will argue
that the original BRST transformations take the 
geometrical form on condition that the matter
currents obey appropriate form of 
the  equations of motion. This
fact however implies that we have to work in the
specific gauge in order to define Hamiltonian
 that determines the time evolution 
in the extended phase space. 

We then show that the non-local charges, defined either
by iteration procedure given in 
  \cite{Brezin:1979am} or by monodromy
matrix are BRST invariant. We define these non-local
charges using the prescription given recently in 
\cite{Hatsuda:2006ts} that allows us to introduce
infinite number of non-local charges even on 
string world-sheet with finite spatial extend.

Using again the prescription given in
\cite{Hatsuda:2006ts} we define the
 monodromy matrix for bosonic string on 
$AdS_5\times S^5$. We again show that 
this monodromy matrix is conserved and
BRST invariant.  

In conclusion, we hope that the BRST treatment of the
bosonic string theory on $AdS_5\times S^5$ background
brings new insight on an
 existence and importance
of the infinite number of non-local and local 
conserved charges. The fact that they are BRST 
invariant emphasize their physical
importance for the description of
given state of the string and consequently
serves as further support of the integrability
of the bosonic string on $AdS_5\times S^5$. 
Of course one can expect that in order to properly
include quantum effects one should take fermions
into account. For that reason 
we  hope that  better 
understanding of the relation between BRST symmetry
and integrability could be useful for further
development of the pure spinor string in 
$AdS_5\times S^5$ background.

This paper is organized as follows. In next
section (\ref{second}) we formulate the bosonic
string on $AdS_5\times S^5$ as  principal model.
We perform its Hamiltonian analysis and determine
all constrains that given theory have to obey.
In section (\ref{third}) we introduce extended
phase space and determine BRST charge and its
action on fundamental fields. Then we perform
 field redefinitions that map the original
BRST transformations to  ones that have
the properties of the geometrical BRST transformations.
In section (\ref{fourth}) we review the
iterative definition of the non-local charges
that holds for general world-sheet metric. 
We will argue that these non-local charges
can be defined on the string world-sheet as well
and that they are conserved. Then we show
that these charges are BRST invariant as well.
Finally in section (\ref{fifth}) we introduce
the monodromy matrix and review its main
properties. We carefully review
 the calculation 
of the Poisson brackets between the
monodromy matrix and any function defined
on extended phase space. Then we will show
that this monodromy matrix has vanishing
Poisson brackets with the Hamiltonian and
with the BRST charge. In other words we show
that the monodromy matrix is time independent
and BRST invariant. 
\section{Bosonic string on $AdS_5\times
S^5$ as principal model}\label{second}
In this section we formulate the bosonic
string on the $AdS_5\times S^5$ as principal
chiral model, following 
 the notation given 
in \cite{Arutyunov:2004yx}. 
 The five sphere $S^5$ is
parameterized by five variables: 
coordinates $y^i \ , i=1,\dots,4$ and
the angle variable $\phi$. In terms of
six real embedding coordinates $Y^A \ , 
A=1,\dots,6$ obeying the condition
$Y_AY^A=1$ the parametrisation reads
\begin{eqnarray}\label{my}
\mY_1=Y_1+iY_2&=&\frac{y_1+iy_2}{
1+\frac{y^2}{4}} \ , \quad 
\mY_2=Y_3+iY_4=\frac{y_3+iy_4}{
1+\frac{y^2}{4}} \ , 
\nonumber \\
\mY_3&=&Y_5+iY_6=\frac{1-\frac{y^2}{4}}{
1+\frac{y^2}{4}}\exp (i\phi)
 \ .
\nonumber \\
\end{eqnarray}
The metric induced on five sphere $S^5$ from
the flat induced metric is
\begin{eqnarray}
dY_AdY_A=
\left(\frac{1-\frac{y^2}{4}}{1+\frac{y^2}{4}}\right)^2d\phi^2+
\frac{1}{(1+\frac{y^2}{4})^2}dy_idy_i \ , 
\nonumber \\
\end{eqnarray}
where $y^2\equiv y_iy_i$. 
In the same way we describe the $AdS_5$
space when  we introduce four coordinates $z_i$ and
$t$. The embedding coordinates $Z_A$ that
obey $Z_AZ_B\eta^{AB}=-1$ with the metric
$\eta^{AB}=(-1,1,1,1,1,-1)$ is now parameterized
as 
\begin{eqnarray} \label{mz}
\mZ_1=Z_1+iZ_2&=&-\frac{z_1+iz_2}{
1-\frac{z^2}{4}} \ , \quad  
\mZ_2=Z_3+iZ_4=-\frac{z_3+iz_4}{
1-\frac{z^2}{4}} \ , 
\nonumber \\
\mZ_3&=&Z_0+iZ_5=\frac{1+\frac{z^2}{4}}{
1-\frac{z^2}{4}}\exp (it)
 \  
\nonumber \\
\end{eqnarray}
so that the induced 
metric takes the form 
\begin{eqnarray}
\eta_{AB}dZ^AdZ^B=
-\frac{(1+\frac{z^2}{4})^2}{(1-\frac{z^2}{4})^2}dt^2
+\frac{1}{(1-\frac{z^2}{4})^2}dz_idz_i \ , 
\nonumber \\
\end{eqnarray}
where again $z^2\equiv z_iz_i$. 
We presume closed string and hence
all fields $Y_A,Z_A$ are periodic functions
of the world-sheet variable $\sigma\in (0,2\pi)$.
It is remarkable fact that the bosonic string
on the $AdS_5\times S^5$ can be formulated as
principal chiral model with the action
\begin{equation}\label{actg}
S=-\frac{\sqrt{\lambda}}{4\pi}
\int d\tau d\sigma \sqrt{-\gamma}
\gamma^{\alpha\beta}\tr (
g^{-1}\partial_\alpha g g^{-1}\partial_\beta
g) \ ,
\end{equation}
where $\sqrt{\lambda}$ is related 
to the radius  $R$ of $S^5$ ($AdS_5$) and 
the slope $\alpha'$ of the Regge trajectory
as  $\sqrt{\lambda}=\frac{R^2}{\alpha'}$ and
where $\gamma^{\alpha\beta} \ , \alpha=\tau,\sigma$
is world-sheet metric. 
In (\ref{actg}) 
the  matrix $g$ takes the form
\begin{equation}
g=\left(\begin{array}{cc}
g_a & 0 \\
0 & g_s \\ 
\end{array}\right) \ . 
\end{equation}
Here $g_a$ and $g_s$
are the following $4\times
4$ matrices
\begin{equation}
g_a=\left(\begin{array}{cccc}
0 & \mZ_3 & -\mZ_2 &\mZ_1^* \\
-\mZ_3& 0 & \mZ_1 & \mZ_2^* \\
\mZ_2 & -\mZ_1 & 0 &-\mZ_3^* \\
-\mZ_1^* & -\mZ_2^* & \mZ_3^*
&0 \\
\end{array}\right) \ , \quad 
g_s=\left(\begin{array}{cccc}
0 & \mY_1 & -\mY_2 &\mY_3^* \\
-\mY_1& 0 & \mY_3 & \mY_2^* \\
\mY_2 & -\mY_3 & 0 & \mY_1^* \\
-\mY_3^* & -\mY_2^* & -\mY_1^*
&0 \\
\end{array}\right) \ ,
\end{equation}
where $\mZ_k, k=1,2,3$ are the
complex embedding coordinates for
$AdS_5$ defined in  (\ref{mz})
 and $\mY_k,k=1,2,3$
are the complex embedding coordinates
for sphere defined in
(\ref{my}).  The matrix
$g_a$ is an element of the group 
$SU(2,2)$ since it can be shown that
\begin{equation}
g_a^\dag E g_a=E \ , \quad
E=\mathrm{diag}(-1,-1,1,1)
\end{equation}
provided the following condition is
satisfied
\begin{equation}
\mZ_1^*\mZ_1+\mZ_2^*\mZ_2
-\mZ_3^*\mZ_3=-1 \ .
\end{equation}
In fact $g_a$ describes embedding
of an element of the coset space
$SO(4,2)/SO(5,1)$ into group 
$SU(2,2)$ that is locally isomorphic to
$SO(4,2)$. We use this isometry to
work with $4\times 4$ matrices
rather with $6\times 6$ ones. Note
that due to the explicit choice of
the coset representative above there
is not any gauge symmetry left.
Quite analogously $g_s$ 
is unitary 
\begin{equation}
g_s g_s^\dag=1
\end{equation}
on condition that $\mY_1^*\mY_1+
\mY_2^*\mY_2+\mY^*_3\mY_3=1$.
The matrix $g_s$ describes 
an embedding of an element 
of the coset $SO(6)/SO(5)$
into $SU(4)$ being isomorphic
to $SO(6)$. In what follows we 
use the abstract formulation 
of the action 
\begin{equation}\label{actpr}
S=-\frac{\sqrt{\lambda}}{4\pi}\int d\tau d\sigma
\sqrt{-\gamma}\gamma^{\alpha\beta}
J^A_\alpha J^B_\beta K_{AB} \ ,  
\end{equation}
where 
\begin{equation}
J=g^{-1}dg=J^AT_A   
\end{equation}
and where $T_A$  are generators of 
the algebra $\mathbf{g}$
with the metric  
\begin{equation}
K_{AB}=\tr (T_AT_B) \ . 
\end{equation}
In order to develop the canonical
formalisms we follow
\cite{Faddeev:1987ph}.
 Using the flatness of the current $J^A$
\begin{equation}
\partial_\tau J^A_\sigma-\partial_\sigma J^A_\tau
+J_\tau^BJ_\sigma^C f_{BC}^A=0
\end{equation}
we can  express $J_\tau^A$ 
as \begin{equation}
\partial_\tau J_\sigma^A=(\partial_\sigma
 \delta^A_C+
J_\sigma^B f_{BC}^A)J_\tau^C\equiv D_C^A J_\tau^C
\end{equation}
that implies 
\begin{equation}\label{J0D}
J_\tau^A=(D^{-1})_B^A\partial_\tau J^B_\sigma \ . 
\end{equation}
We see that it is natural to interpret
$J_\sigma^A$ as a  canonical variable with the
corresponding 
conjugate momentum $\Pi_A$. These
canonical variables have following
 Poisson bracket
\begin{equation}\label{canpb}
\pb{J_\sigma^A(\sigma_1),\Pi_B(\sigma_2)}=\delta^A_B
\delta(\sigma_1-\sigma_2) \ . 
\end{equation}
In order to find the corresponding 
 momentum $\Pi_A$ we
insert  (\ref{J0D}) into the action
(\ref{actpr}) and perform the variation 
with respect to $\partial_\tau J^A_\sigma$.
After some calculations we obtain
\begin{equation}
\Pi_A=\frac{\delta S}{\delta 
\partial_\tau J_\sigma^A}= \frac{\sqrt{\lambda}}{2\pi}
K_{AB}(D^{-1})^B_C(
\sqrt{-\gamma}\gamma^{\tau\tau}(D^{-1})^C_D
(\partial_\tau J_\sigma^D)+\sqrt{-\gamma}
\gamma^{\tau\sigma}
J_\sigma ^C) \ . 
\end{equation}
Using this relation we can express 
$J_\tau^A$ as a function of $J^A_\sigma$ and
$\Pi_A$
\begin{eqnarray}\label{j0ap}
 J_\tau^A=\frac{1}{\sqrt{-\gamma}\gamma^{\tau\tau}}
 (\frac{2\pi}{\sqrt{\lambda}}
\mD^A-\sqrt{-\gamma}\gamma^{\tau\sigma}
 J^A_\sigma) \ ,  \nonumber \\
\end{eqnarray}
where we have defined
\begin{equation}
\mD^A\equiv D^A_B\Pi^B \ . 
\end{equation}
As the next step we define the Hamiltonian as
\begin{eqnarray}\label{H0}
H_0&=&\int d\sigma \mH_0=
\int d\sigma (\Pi_A \partial_\tau J_\sigma^A-\mL)=
\nonumber \\
&=&\int d\sigma 
\left[\frac{\sqrt{-\gamma}}{\gamma_{\sigma\sigma}}(
\frac{\pi}{\sqrt{\lambda}}\mD^C\mD^DK_{CD}
+\frac{\sqrt{\lambda}}{4\pi}J^C_\sigma J^D_\sigma K_{CD})+
\frac{\gamma_{\tau\sigma}}
{\gamma_{\sigma\sigma}}J^C_\sigma K_{CD} \mD^D \right]\ . 
\nonumber \\
 \end{eqnarray}
We choose the 
parametrisation of the metric
 $\gamma_{\alpha\beta}$ as 
 \cite{Fujiwara:1992eb}
\begin{equation}\label{pame}
\lambda^\pm=\frac{\sqrt{-\gamma}
\pm \gamma_{\tau\sigma}}{\gamma_{\sigma\sigma}} \ , 
\quad \xi=\ln \gamma_{\sigma\sigma} \ ,
\end{equation}
where $\lambda^\pm$ are manifestly
invariant under Weyl transformation
\begin{equation}
g'_{\alpha\beta}(\sigma,\tau)=
e^{\phi(\sigma,\tau)}g_{\alpha\beta}
(\sigma,\tau) \  
\end{equation}
 while
$\xi$ transforms as $\xi'(\sigma,\tau)=
\xi(\sigma,\tau)+\phi(\sigma,\tau)$. 
Since the action (\ref{actpr}) does
not contain time-derivative of $\gamma_{\alpha\beta}$
it follows that the momenta conjugate
to $\lambda^\pm,\xi$ are zero:
\begin{equation}\label{pim}
\pi^\lambda_\pm=\frac{\delta S}{\delta 
\partial_\tau\lambda^\pm}=0 \ , \quad
\pi_\xi=\frac{\delta S}{\delta
\partial_\tau \xi}=0 \ . 
\end{equation}
These conditions consist primary constraints
of the theory. Using (\ref{pame})
the Hamiltonian
density  (\ref{H0}) takes the form
\begin{eqnarray}
\mH_0&=&
\frac{\sqrt{-\gamma}}{\gamma_{\sigma\sigma}}(
\frac{\pi}{\sqrt{\lambda}}\mD^C\mD^DK_{CD}
+\frac{\sqrt{\lambda}}{4\pi}J^C_\sigma
 J^D_\sigma K_{CD})+
\frac{\gamma_{\tau\sigma}}
{\gamma_{\sigma\sigma}}J^C_\sigma K_{CD} \mD^D
=\nonumber \\
&=&
\frac{\sqrt{-\gamma}}{\gamma_{\sigma\sigma}}T_0+
\frac{\gamma_{\tau\sigma}}{\gamma_{\sigma\sigma}}
T_1=\frac{\lambda^+}{2}(T_0+T_1)+
\frac{\lambda^-}{2}(T_0-T_1) \ .
\nonumber \\
\end{eqnarray}
We  see an advantage of the 
parametrisation (\ref{pame}) since the
Hamiltonian density $\mH_0$ 
 contains   variables $\lambda^{\pm}$ only
and hence it is Weyl invariant.
According to the general analysis the
time evolution of the primary constraints
$\pi^\lambda_{\pm}$ imply an existence of the
secondary constraints 
\begin{equation}\label{Tpm}
T_\pm=\frac{1}{2}(T_0\pm T_1)\approx 0 \ . 
\end{equation}
As the next step we determine
some fundamental Poisson brackets.
Recall that $\mD^A$ is equal 
to 
\begin{equation}
\mD^A=
\partial_\sigma \Pi^A+J_\sigma^B \Pi^C
f_{BC}^A \ . 
\end{equation}
Then using the canonical Poisson
bracket (\ref{canpb})
it is easy to see that
\begin{eqnarray}
\pb{J_\sigma^A(\sigma),\mD^B(\sigma')}&=&
-\partial_\sigma
\delta(\sigma-\sigma')K^{AB}-
J^C_\sigma(\sigma)f_{CD}^A K^{DB}
\delta(\sigma-\sigma')
 \ , 
\nonumber \\
\pb{\Pi^A(\sigma),\mD^B(\sigma')}&=&
-K^{AC}f_{CD}^B\Pi^D(\sigma)\delta(\sigma-\sigma') \ ,
\nonumber \\
\pb{\mD^A(\sigma),\mD^B(\sigma')}&=&
-K^{AC}f_{CD}^B \mD^D(\sigma)\delta(\sigma-\sigma')  \ . 
\nonumber \\
\end{eqnarray}
As the next step we determine the Poisson
brackets between $J_\sigma,\mD$ and $T_{0,1}$
\begin{eqnarray}\label{j10t}
\pb{J_\sigma^A(\sigma),T_0(\sigma')}&=&
\partial_{\sigma'}\delta(\sigma-\sigma')\sqrt{-\gamma}
\gamma^{\tau\alpha}J_\alpha^A(\sigma')-
\sqrt{-\gamma}\gamma^{\tau\tau}
J^B_\sigma J^C_\tau(\sigma') f_{BC}^A
\delta(\sigma-\sigma') \ ,
\nonumber \\
\pb{J_\sigma^A(\sigma),T_1(\sigma')}&=&
\partial_{\sigma'} 
\delta(\sigma-\sigma')J^A_\sigma(\sigma') \ , 
 \nonumber \\
\pb{\mD^A(\sigma),T_1(\sigma')}&=&
\partial_{\sigma'}\delta(\sigma-\sigma')
\mD^A(\sigma') \ , 
\nonumber \\
\pb{\mD^A(\sigma),T_0(\sigma')}&=&
\frac{\sqrt{\lambda}}{2\pi}
\partial_{\sigma'} \delta(\sigma-\sigma')
J^A_\sigma(\sigma') \ . 
\nonumber \\
\end{eqnarray}
Using these Poisson brackets we
  easily determine  
\begin{eqnarray}
\pb{T_1(\sigma),T_1(\sigma')}
&=&-2T_1(\sigma)\partial_\sigma
\delta(\sigma-\sigma')-
\partial_\sigma T_1(\sigma)
\delta(\sigma-\sigma') \ ,
\nonumber \\
\pb{T_0(\sigma),T_1(\sigma')}&=&
-2T_0(\sigma)\partial_\sigma
\delta(\sigma-\sigma')-
\partial_\sigma T_0(\sigma)
\delta(\sigma-\sigma') \ , 
\nonumber \\
\pb{T_0(\sigma),T_0(\sigma')}&=&
-2T_1(\sigma)
\partial_\sigma\delta(\sigma-\sigma')-
\partial_\sigma T_1(\sigma)
\delta(\sigma-\sigma') \ . 
\nonumber \\
\end{eqnarray}
It is convenient to define  
following combinations
\begin{equation}
T_+=\frac{1}{2}
(T_0+T_1) \ , \quad  T_-=\frac{1}{2}
(T_0-T_1) \ 
\end{equation}
that have following Poisson
brackets 
\begin{eqnarray}
\pb{T_+(\sigma),T_+(\sigma')}&=&
-2T_+(\sigma)\partial_\sigma
\delta(\sigma-\sigma')
-\partial_\sigma 
T_+(\sigma)\delta(\sigma-\sigma') \ , 
\nonumber \\
\pb{T_-(\sigma),T_-(\sigma')}&=&
2T_-(\sigma)\delta(\sigma-\sigma')+
\partial_\sigma T_-(\sigma)
\delta(\sigma-\sigma') \ , \nonumber \\
\pb{T_+(\sigma),T_-(\sigma')}&=& 0 \ . 
\nonumber \\
\end{eqnarray}
After determination of 
 these Poisson brackets we
can proceed to the BRST analysis
of the principal model.
\section{Extended phase space for
 principal model}\label{third}
 We introduce the extended phase space
 following the approach presented in 
 \cite{Fujiwara:1992eb}.
The extended phase space of the BFV
theory is defined as including to the
classical phase space the ghost-auxiliary
field sector
\begin{equation}
(\mC^A,\omP_A)  \ ,
(\mP^A,\omC_A) \ , 
(N^A,B_A) \ ,  
\end{equation}
where $A=(\lambda^\pm,\xi,\pm)$ label the
first class constraints $\pi^\lambda_\pm=
\pi_\xi=0$ and $T_\pm=0$. $\mC^A$ and
$\mP^A$ are the BFV ghosts fields carrying one
unit of the ghost number while
$\omP_A$ and $\omC_A$ are their canonical momenta.
The last canonical pairs are auxiliary fields that 
carry zero ghost number. 
Note that these fields have following Poisson
brackets
\begin{eqnarray}\label{ghospb}
\pb{\mC^A(\sigma),\omP_B(\sigma')}&=&
\pb{\omP_B(\sigma'),\mC^A(\sigma)}=
-\delta^A_B\delta(\sigma-\sigma') \ , 
\nonumber \\
\pb{\mP^A(\sigma),\omC_B(\sigma')}&=&
\pb{\omC_B(\sigma'),\mP^A(\sigma)}=
-\delta^A_B\delta(\sigma-\sigma') \ , 
\nonumber \\
\pb{N^A(\sigma),B_B(\sigma')}&=&
-\pb{B_B(\sigma'),N^A(\sigma)}=
\delta^A_B\delta(\sigma-\sigma') \ . 
\nonumber \\
\end{eqnarray}
Given the constraints (\ref{pim})
and (\ref{Tpm}) we can construct the BRST
charge in standard way
\begin{eqnarray}\label{Q}
Q&=&\int d\sigma
[\mC^+_\lambda \pi_+^\lambda+
\mC^-_\lambda \pi_-^\lambda+
\mC^\xi \pi_\xi +
\mC^+(T_++\omP_+\partial_\sigma \mC^+)+
\nonumber \\
&+&\mC^-(T_--\omP_-\partial_\sigma \mC^-)
+B_A\mP^A] \ .
\nonumber \\
\end{eqnarray}
Using the Poisson brackets given
in (\ref{ghospb}) it is 
straightforward to determine 
the BRST transformations of the
ghost fields 
\begin{eqnarray}
\delta \lambda^\pm&=&
\pb{\lambda^\pm,Q}=\mC^\pm_\lambda \ , 
\nonumber \\
\delta \xi&=&\pb{\xi,Q}=\mC^\xi \ , 
\nonumber \\
\delta \pi_\pm^\lambda&=&\delta \pi_\xi=0 \ , 
\quad \delta \mC^\lambda_\pm=\delta C^\xi=0 \ ,
\nonumber \\
\delta \omP^\lambda_\pm&=&
\pb{\omP^\lambda_\pm,Q}=
-\pi^\lambda_\pm \ , \nonumber \\
\delta \omP_\xi&=&
\pb{\omP_\xi,Q}=-\pi_\xi \ ,
\nonumber \\
\delta \mP^\pm &=&0 \ , \quad \delta N^A=
\pb{N^A,Q}=-\mP^A \ , 
\nonumber \\
\delta\omC_A&=& \pb{\omC_A,Q}=
-B_A \ , \quad  \delta B_A=0 \ . 
\nonumber \\
\end{eqnarray}
In order to determine the BRST
transformation of the current
$J^A_{\sigma}$ we use the Poisson brackets
(\ref{j10t}) together with 
(\ref{ghospb}) and we get
\begin{eqnarray}\label{jsigmaQ} 
\delta_Q J^A_\sigma(\sigma)&=&
\pb{J^A_\sigma(\sigma),Q}=
-\frac{(\mC^++\mC^-)}{2}
(\sqrt{-\gamma}\gamma^{\tau\sigma}\partial_\sigma J_\sigma^A+
\sqrt{-\gamma}\gamma^{\tau\tau}\partial_\sigma J_\tau^A)-
\nonumber \\
&-&
\frac{(\mC^+-\mC^-)}{2}
\partial_\sigma J_\sigma^A-
\frac{\partial_\sigma [(\mC^++\mC^-)
\sqrt{-\gamma}\gamma^{\tau \alpha}]}{2}
J_\alpha^A
-\frac{\partial_\sigma(\mC^+-\mC^-)}{2}
J_\sigma^A  \ , \nonumber \\
\end{eqnarray}
where $\alpha=\tau,\sigma$. In
principle the metric components $\gamma_{\alpha\beta}$
can be expressed using the variables
$\lambda^\pm, \xi$ but it will not be
necessary. 
In order to find the 
BRST transformation of $J_\tau^A$
we have to use the relation 
(\ref{j0ap}) and then the
Poisson brackets given in
 (\ref{j10t}),(\ref{ghospb}). After
 some calculations we obtain
\begin{eqnarray}\label{jtauQ}
\delta_QJ^A_\tau=
\pb{J_\tau^A(\sigma),Q}&=&
\frac{(\mC^++\mC^-)}{2}
(\sqrt{-\gamma}\gamma^{\tau\sigma}
\partial_\sigma J_\sigma^A+
\sqrt{-\gamma}\gamma^{\tau\tau}\partial_\tau J_\sigma^A)
-\frac{1}{2\sqrt{-\gamma}
\gamma^{\tau\tau}}\partial_\sigma(\mC^++\mC^-)J_\sigma^A
+\nonumber \\
&+&\frac{\gamma^{\tau\sigma}}{\gamma^{\sigma\sigma}}
\partial_\sigma[(\mC^++\mC^-)\sqrt{-\gamma}\gamma^{\tau\alpha}]
J_\alpha^A-
\frac{1}{2}(\mC^+-\mC^-)\partial_\sigma J_\tau^A
-\nonumber \\
&-&\frac{1}{2\sqrt{-\gamma}\gamma^{\tau\tau}}
\partial_\sigma[(\mC^+-\mC^-)\sqrt{-\gamma}
\gamma^{\tau\alpha}]J^A_\alpha+
\frac{\gamma^{\tau\sigma}}{2\gamma^{\tau\tau}}
\partial_\sigma(\mC^+-\mC^-)J_\sigma^A-
\nonumber \\
&-&\frac{1}{2}(\mC^+_\lambda+\mC^-_\lambda)
\frac{2\pi}{\sqrt{\lambda}}
\mD^A
+\frac{(\mC^+_\lambda-\mC^-_\lambda)}
{2}J_\sigma^A\ . \nonumber \\
\end{eqnarray}
We see that the BRST transformations
expressed using the BFV ghosts is
rather complicated.
 On the other hand the Lax
connection and monodromy matrix are
defined using the covariant currents
$J_{\alpha}$ so that it was
necessary to find
the BRST transformation
of $J_\tau$. However we will
argue below that it is possible
to express the BRST transformation
of the currents in more geometrical
form.

Before we proceed to the study of 
this question we focus  our 
attention on
 gauge fixed action for
fields defined in 
the extended phase space. 
The gauge fixed action in the
BFV formalism is defined as
\begin{eqnarray}
S&=&\int d\tau d\sigma
[\dot{J}^A\Pi_A+
\dot{\mC}^A\omP_A+\dot{\lambda}^+\pi^\lambda_++
\dot{\lambda}^-\pi^\lambda_-
+\dot{\lambda}^\xi
\pi_\xi
-\mH]=\nonumber \\
&=&
\int d\tau d\sigma
[\frac{2}{\lambda^++\lambda^-}
\frac{\sqrt{\lambda}}{2\pi}J_\tau^AK_{AB}J^B_\tau
+\frac{\lambda^+-\lambda^-}
{\lambda^++\lambda^-}
\frac{\lambda}{2\pi}
J_\tau^AK_{AB}J^B_\sigma+\nonumber \\
&+&\dot{\mC}^A\omP_A
+\dot{\lambda}^+\pi^\lambda_++
\dot{\lambda}^-\pi_-^\lambda+\dot{\lambda}^\xi
\pi_\xi-\mH] \ , 
\nonumber \\
\end{eqnarray}
where $\dot{X}\equiv \partial_\tau (X)$ and where
\begin{equation}
H\equiv \int d\sigma\mH=
H_0-\pb{Q,\Psi} \ . 
\end{equation}
 $\Psi$  given above is the gauge
fermion.
Note that we have used the fact that
\begin{equation}
\omC_A\dot{\mP}^A+B_A\dot{N}^A=
-\pb{\omC_A\dot{N}^A,Q}
\end{equation}
and hence this term can be absorbed
into $\Psi$ by redefinition of $\chi^A$. 

As we have seen 
$\mH_0$ is a sum of the constraints
$T_\pm$. Then according to the general
principles of the BRST Hamiltonian
formalism we say that $\mH_0=0$.

  Clearly
we can not go further without any
assumption on $\Psi$ so that we consider
it in the standard form
\begin{equation}
\Psi=\int d\sigma
[\omC_A\chi^A+\omP_A N^A] \ , 
\end{equation} 
where $\chi_A$ are gauge-fixing functions. 
With this choice of $\Psi$ we obtain
the Hamiltonian in the form
\begin{eqnarray}
H=-\pb{\Psi,Q}&=&\int d\sigma (\mH_{CL}+
\mH_{GF}+\mH_{FP}) \ , 
\nonumber \\
\mH_{CL}&=&
N^+T_++N^-T_-
+\pi^\lambda_+N^+_\lambda
+\pi_-^\lambda N^-_\lambda+
\pi_\xi N^\xi=
\nonumber \\
&=&\frac{1}{2}(N^++N^-)T_0+
\frac{1}{2}(N^+-N^-)T_1+
+\pi^\lambda_+N^+_\lambda+
\pi_-^\lambda N^-_\lambda+
\pi_\xi N^\xi
 \ , \nonumber \\
\mH_{GF}&=&B_A\chi^A \ , \nonumber \\
\mH_{FP}&=&
-\omC_A\pb{\chi^A,Q}+\omP_A\mP^A+
\nonumber \\
&+&[2\omP_+\partial_\sigma \mC^++\partial_\sigma
\omP_+ \mC^+ ]N^-
-[2\omP_-\partial_\sigma\mC^-+\partial_\sigma
\omP_-\mC^-]N^- \ . 
\nonumber \\
\end{eqnarray}
Comparing $\mH_C$ with $\mH_0$ 
suggests that  $N^\pm$
can be related to two of the metric
variables as 
\begin{eqnarray}
N^0&=&\frac{1}{2}(N^++N^-)=
\frac{1}{\sqrt{-\gamma}\gamma^{\tau\tau}}=
\frac{1}{2}(\lambda^++\lambda^-)
 \ , \nonumber \\
N^1&=&\frac{1}{2}(N^+-N^-)=\frac{
\gamma_{\tau\sigma}}{\gamma_{\sigma\sigma}}
=\frac{1}{2}(\lambda^+-\lambda^-) \ . 
\nonumber \\
\end{eqnarray}
Then in order to
 recover the original form of the metric
we impose following gauge conditions
\begin{equation}\label{chigf}
\chi^+_\lambda=
\lambda^+-N^+ \ , \quad 
\chi^-_\lambda=\lambda^--N^-
\end{equation}
without changing the relation 
$\xi=\ln g_{11}$. Then the solving
the equation of motion with respect
to $B_{+,-}^\lambda$ we obtain
that 
\begin{equation}
\chi^+_\lambda=0 \ , \quad 
\chi^-_\lambda=0 \ . 
\end{equation}
On the other hand if we calculate
the Poisson brackets between
$\chi^\pm_\lambda$ given in 
(\ref{chigf}) and $Q$  we obtain
\begin{equation}
-\pb{\chi^+_\lambda,Q}=
-\mC^+_\lambda+\mP^+ \ , \quad 
-\pb{\chi^-_\lambda,Q}=
-\mC^-_\lambda+\mP^- \ . 
\end{equation}
Then the gauge fixed action takes
the form (after integration out
$B^\lambda_{\pm}$)
\begin{eqnarray}\label{Sgafix}
S&=&S_{matter}+S_{ghosts}+S_{FG} \ , 
\nonumber \\ 
S_{matter}&=&
\int d\sigma d\tau 
[\dot{\lambda}^+\pi^\lambda_+
+\dot{\lambda}^-\pi^\lambda_-
+\dot{\lambda}^\xi\pi_\xi-\nonumber \\
&-&\frac{\sqrt{\lambda}}{4\pi}
\sqrt{-\gamma}\gamma^{\alpha\beta}
J_\alpha^K J_\beta^LK_{KL}-
\pi^\lambda_+N^+_\lambda-
\pi_-^\lambda N^-_\lambda-
\pi_\xi N^\xi_\lambda] \ , 
\nonumber \\
S_{ghosts}&=&
\int d\sigma d\tau[
\dot{\mC}^A\omP_A-\mH_{FP}] \ , 
\nonumber \\
\mH_{FP}&=&-\omC_+\pb{\chi^+,Q}-
\omC_-\pb{\chi^-,Q}-\omC_\xi
\pb{\chi^\xi,Q}-\nonumber \\
&-&\omC_{+}^\lambda \mC^+_\lambda
+\omC_{+}^\lambda\mP^+
-\omC_{-}^\lambda \mC^-_\lambda
+\omC_-^\lambda \mP^-
+\omP_A\mP^A+
\nonumber \\
&+&[2\omP_+\partial_\sigma \mC^++\partial_\sigma
\omP_+ \mC^+ ]\lambda^+
-[2\omP_-\partial_\sigma\mC^-+\partial_\sigma
\omP_-\mC^-]\lambda^-
 \ , \nonumber \\
S_{FG}&=&-\int d\sigma d\tau
(B_+\chi^++B_-\chi^-+B_\xi \chi^\xi)
 \ , \nonumber \\
\nonumber \\
\end{eqnarray}
where we have used 
\begin{eqnarray}& &
\frac{1}{\sqrt{-\gamma}\gamma^{\tau\tau}}
T_0(x)+\frac{\gamma^{\tau \sigma}}
{\gamma^{\sigma\sigma}}T_1(x)=
\nonumber \\
&=&
\frac{\sqrt{\lambda}\sqrt{-\gamma}}{4\pi}
(\gamma^{\tau\tau}J_\tau^AJ_\tau^BK_{AB}-\gamma^{\sigma\sigma}
 J_\sigma^A J_\sigma^BK_{AB}) \nonumber \\
\end{eqnarray}
and the relation (\ref{j0ap}). 
After introduction of the
basic object of the BFV formalism
we  come to the question how
to simplify the BRST transformations
of the currents. It turns out that
the natural way how to do this is
to introduce  more familiar 
geometrical BRST transformations. 
In fact, the primary constraints $\pi^\lambda_\pm$
and $\pi_\xi$ are responsible for the
fact that the variables $\lambda^\pm$ and
$\xi$ have lost their original meanings.

\subsection{Review of the geometrical
BRST transformations}
In this subsection we review
 the geometrical BRST transformations.
These BRST transformations have their
origin in the  transformations of the world-sheet
modes  under the diffeomorphism and
Weyl transformations. 

Let us 
consider following diffeomorphism 
transformations
\begin{equation}
x'^\mu=x^\mu+\omega^\mu(x) \ ,
\quad \frac{\partial x'^\alpha}{\partial x^\mu}
\equiv \partial_\mu x'^\alpha=
\delta_\mu^\alpha+\partial_\mu\omega^\alpha \ ,    
\end{equation}
where  $x^\mu\equiv(x^0=\tau,
x^1=\sigma)$. 
Since  currents are 
one-form on the world-sheet
their  transformation properties
can be easily determined from 
\begin{eqnarray}
J'_\mu(x')dx'^\mu&=&J_\mu (x)dx^\mu
\Rightarrow
J'_\mu(x')=J_\alpha \left(\frac{\partial
x'^\mu}{\partial x^\alpha}
\right)^{-1} \Rightarrow \nonumber \\
\Rightarrow 
\delta J_\mu(x)&=&J'_\mu(x)-J_\mu(x)=
-\partial_\alpha J_\mu \omega^\alpha
-J_\alpha(x)\partial_\mu \omega^\alpha \ .
\nonumber \\
\end{eqnarray}
Analogously we determine the
variation of the metric
under diffeomorphism transformations
\begin{equation}
\delta \gamma_{\mu\nu}=
-\partial_\alpha \gamma_{\mu\nu}\omega^\alpha
-\gamma_{\mu\alpha}\partial_\nu \omega^\alpha-
\partial_\mu \omega^\alpha \gamma_{\alpha \nu}
\end{equation}
while its variation under
 Weyl transformation takes the form
\begin{equation}
\delta_W \gamma_{\alpha\beta}=-\omega_W \gamma_{\alpha\beta} \ . 
\end{equation}
For letter purposes we also determine
the  variation of $\gamma_{\alpha\beta}
\epsilon^{\beta\gamma}$ where 
\begin{equation}
\epsilon^{\mu\nu}=\frac{e^{\mu\nu}}{
\sqrt{-\gamma}} \ , \quad 
e^{01}=-e^{10}=1 \ . 
\end{equation}
Since $(\gamma\epsilon)^\gamma_\alpha$
is an object with one lower and one upper
index it is clear that its 
variation takes the form 
\begin{equation}
\delta (\gamma_{\alpha\beta}
\epsilon^{\beta\gamma})=
-\omega^\delta\partial_\delta (\gamma_{\alpha\beta}
\epsilon^{\beta\gamma})
-\partial_\alpha \omega^\delta 
\gamma_{\delta\beta}\epsilon^{\beta\gamma}
+\gamma_{\alpha\beta}\epsilon^{\beta\delta}
\partial_\delta \omega^\gamma
\end{equation}
while it is invariant under Weyl transformations. 
Finally, using the fact that
\begin{equation}
\sqrt{-\gamma'(x')}=\sqrt{-\gamma(x)}
|\det (\partial_\alpha x'^\mu)^{-1}|
\end{equation}
we get
\begin{equation}
\delta \sqrt{-\gamma(x)}=
-\partial_\alpha[\sqrt{-\gamma}]\omega^\alpha
-\partial_\alpha \omega^\alpha \sqrt{-\gamma}
\end{equation}
while its variation under Weyl
transformation takes the form
\begin{equation}
\delta_W \sqrt{-\gamma}=\omega_W
\sqrt{-\gamma} \ . 
\end{equation}
These transformations appear as the BRST 
transformations in the extended
 configuration space. The extended
configuration space contains
 three independent metric
components $\gamma_{\alpha\beta}$,
 currents $J_\alpha$ with two reparametrisation
 ghosts $C^\alpha$ 
and corresponding anti-ghosts
$\overline{C}_\alpha$, one Weyl-ghost $C_W$ and
anti-ghost $\overline{C}_W$. The covariant
BRST transformations follow from the 
variations given above where 
gauge parameters are replaced with corresponding
ghosts. Explicitly, we obtain
\begin{eqnarray}\label{deltaG}
\delta_{Q}
\gamma_{\mu\nu}&=&
\partial_\alpha \gamma_{\mu\nu}C^\alpha
+\gamma_{\mu\alpha}\partial_\nu C^\alpha+
\partial_\mu C^\alpha \gamma_{\alpha \nu}
+C_W \gamma_{\mu\nu} \ , 
\nonumber \\
\delta_{Q} (\gamma_{\alpha\beta}
\epsilon^{\beta\gamma})&=&
C^\delta\partial_\delta (\gamma_{\alpha\beta}
\epsilon^{\beta\gamma})
+\partial_\alpha C^\delta 
\gamma_{\delta\beta}\epsilon^{\beta\gamma}
-\gamma_{\alpha\beta}\epsilon^{\beta\delta}
\partial_\delta C^\gamma \ , 
\nonumber \\
\delta_{Q} J_\mu&=&
\partial_\alpha J_\mu C^\alpha
+J_\alpha\partial_\mu C^\alpha  \ ,  
\nonumber \\
\delta_{Q} J^\mu&=&
\partial_\alpha J^\mu C^\alpha -
J^\alpha \partial_\alpha \epsilon^\mu  \ , 
\nonumber \\
\delta_Q C^\alpha&=& C^\beta\partial_\beta C^\alpha \ ,
\nonumber \\
\delta_Q C_W&=&C^\alpha \partial_\alpha C_W \ , 
\nonumber \\
\delta_{Q} \sqrt{-\gamma}&=&
\partial_\alpha[\sqrt{-\gamma}]C^\alpha
+\partial_\alpha C^\alpha \sqrt{-\gamma} \ . 
\end{eqnarray}
Let us now return to the principal chiral model 
and consider the Lax connection
\begin{equation}\label{Lax}
\hJ_\alpha^A(\Lambda)=\frac{1}{1-\Lambda^2}
(J_\alpha^A-\Lambda  \gamma_{\alpha\beta}
\epsilon^{\beta\gamma}J_\gamma^A) \ , 
\end{equation}
where $\Lambda$ is spectral parameter. 
Using (\ref{deltaG}) it is easy to 
determine the BRST transformation of
the flat connection (\ref{Lax})
\begin{eqnarray}
\delta_{Q}\hJ^A_\alpha
&=&C^\beta\partial_\beta[\frac{1}{1-\Lambda^2}(
J_\alpha^A-\Lambda \gamma_{\alpha\gamma}
\epsilon^{\gamma\delta}J^A_\delta)]+
\frac{1}{1-\Lambda^2}
J^A_\beta \partial_\alpha C^\beta-
\nonumber \\
&-&\frac{\Lambda}{1-\Lambda^2}[(
\partial_\alpha C^\delta \gamma_{\delta\beta}
\epsilon^{\beta\gamma}J^A_\gamma
-\gamma_{\alpha\beta}\epsilon^{\beta\delta}
\partial_\delta  C^\gamma J^A_\gamma)
-\gamma_{\alpha\beta}\epsilon^{\beta\gamma}
(\partial_\gamma C^\delta J_\delta^A)]= 
\nonumber \\
&=&\partial_\alpha C^\beta \hJ^A_\beta+
\partial_\beta \hJ_\alpha^A C^\beta \ .
\nonumber \\
\end{eqnarray}
For letter purposes we also determine
following BRST variations
\begin{eqnarray}\label{heBRST}
\delta_{Q}(\sqrt{-\gamma}\gamma^{\tau\alpha}J_\alpha)
&=&\partial_\tau (\sqrt{-\gamma}J^\tau)C^\tau+
\partial_\sigma (\sqrt{-\gamma}J^\tau C^\sigma)
-\sqrt{-\gamma} J^\sigma \partial_\sigma C^\tau \ , 
\nonumber \\
\delta_{Q}(\sqrt{-\gamma}
\gamma^{\sigma\alpha}J_\alpha)
&=&\partial_\sigma(\sqrt{-\gamma} J^\sigma)C^\sigma+
\partial_\tau (\sqrt{-\gamma} J^\sigma C^\tau)-
\sqrt{-\gamma}J^\tau\partial_\tau C^\sigma \ . 
\nonumber \\
\end{eqnarray}
After this review of the geometrical
BRST transformations we return to the question how
the BFV ghosts are related to the geometrical ghosts
$C^\alpha,C_W$. This problem was carefully
studied in \cite{Fujiwara:1992eb}. It was shown here
that in order to find such a relation we have
to use the equations of motion for BFV ghosts and
consequently we have to specify remaining
functions $\chi^{\pm},\chi^\xi$ in the gauge
fermion. It was argued that it is sufficient to presume
that  these functions do not depend on 
$\omP_A,\pi_{\pm}^\lambda$ and $\pi_\xi$ but are 
arbitrary otherwise.  Then it was shown 
\cite{Fujiwara:1992eb}
that the geometrical
ghosts are related to the BFV ghosts as
\footnote{It was shown in 
\cite{Fujiwara:1992eb} that in the same
way we can find the relation between $C_W$ and
BFV ghosts. Since these formulas will not be
important in what follows we do not include
them here.}
\begin{eqnarray}\label{c0}
C^\tau=\frac{\mC^\tau}{\lambda^\tau} \ , \quad 
C^\sigma=-\mC^\sigma+
\frac{\lambda^\sigma}{\lambda^\tau}\mC^\tau 
\ , \nonumber \\
\end{eqnarray}
where $\mC^\pm=\mC^\tau\pm \mC^\sigma$. 
In order to express the BRST
transformations (\ref{jsigmaQ}) and
(\ref{jtauQ}) using the ghosts
$C^\pm$ we have to 
 invert the relation 
(\ref{c0}) and we get 
\begin{eqnarray}\label{c0i}
\mC^\tau&=&\frac{1}{2}(\mC^++\mC^-)=\lambda^\tau C^\tau=
\frac{\sqrt{-\gamma}}{\gamma_{\sigma\sigma}}C^\tau \ , 
\nonumber \\
\mC^\sigma &=&\frac{1}{2}(\mC^+-\mC^-)=
-C^\sigma
+\frac{1}{2}(\lambda^+-\lambda^-)C^\tau=
-C^\sigma+\frac{\gamma^{\tau\sigma}}
{\gamma^{\tau\tau}}C^\tau \ . 
\nonumber \\
\end{eqnarray}
With the help of these results we obtain
that (\ref{jsigmaQ}) takes the form 
\begin{eqnarray}
\pb{J^A_\sigma(\sigma),Q}=
C^\tau\partial_\tau J_\sigma^A+
C^\sigma\partial_\sigma J_\sigma^A+
\partial_\sigma C^\tau J_\tau^A+\partial_\sigma 
C^\sigma J_\sigma^A
\nonumber \\
\end{eqnarray}
that coincides with the variation
of $J_\sigma$ given in (\ref{deltaG}).
On the other hand if
we use (\ref{c0i}) in (\ref{jtauQ}) we
obtain more complicated result
\begin{eqnarray}\label{deltaJtau}
\pb{J_\tau^A,Q}=
\partial_\tau C^\tau J_\tau^A+\partial_\tau
 C^\sigma J^A_\sigma+
+C^\sigma\partial_\sigma J_\tau^A
+\nonumber \\
+C^\tau[-\frac{1}{\sqrt{-\gamma}\gamma^{\tau\tau}}
\partial_\tau[\sqrt{-\gamma}\gamma^{\tau\sigma}J_\sigma]
-\frac{1}{
\sqrt{-\gamma}\gamma^{\tau\tau}}
\partial_\sigma[\sqrt{-\gamma}\gamma^{\sigma \tau}J_\tau^A]-
\nonumber \\
-\frac{1}{\gamma^{\tau\tau}
\sqrt{-\gamma}}\partial_\sigma[\sqrt{-\gamma}
\gamma^{\sigma\sigma}J_\sigma]-\frac{1}{\sqrt{-\gamma}
\gamma^{\tau\tau}}\partial_\tau
[\sqrt{-\gamma}\gamma^{\tau\tau}]
J_\tau^A] \ , \nonumber \\
\end{eqnarray}
where we have also used the equations
of motion for  $\mC^\pm_\lambda$ in order to
express them as functions of $\mC^\pm$ and
 $\lambda^\pm$.
We see that this expression can be
written in the  standard form
 on condition that
currents obey the equation of motion
\begin{equation}\label{eqm}
\partial_\alpha[\sqrt{-\gamma}
\gamma^{\alpha\beta}J_\beta^A]=0\ 
\end{equation}
that follow from the variation of the
action (\ref{Sgafix}) on condition that
the gauge fixing functions 
$\chi^\pm,\chi^\xi$ do not depend on 
$J$. Then  if we 
demand that currents obey (\ref{eqm})
we can rewrite (\ref{deltaJtau})
into the form
\begin{equation}
\pb{J_\tau^A,Q}=
C^\alpha \partial_\alpha J_\tau^A+
J^A_\alpha \partial_\tau C^\alpha 
\end{equation}
that  coincides with the variation
of $J_\tau$ 
given in (\ref{deltaG}). 
\section{BRST invariance of the
non-local conserved charges}
\label{fourth}
\subsection{Definition of the 
non-local charges for general
world-sheet metric}
It is easy to see that
the currens $J=g^{-1}dg$ and
consequently  the action
(\ref{Sgafix}) are invariant 
under the left multiplication of
the group element $g$ by constant
matrix $h$ from $G$. 
Then for $h=1+\epsilon$, where
$\epsilon=\epsilon^AT_A$
 belongs to $\mathbf{g}$
the algebra of $G$ the
variation of the current is
equal to
\begin{equation}
\delta J=g^{-1}d\epsilon g 
\end{equation}
or in components 
($\delta J=\delta J^AT_A \ , 
\epsilon=\epsilon^AT_A$)
\begin{equation}
\delta J^A=d\epsilon^B C_{BC}K^{CA} \  , \quad  
C_{BC}=\tr (g^{-1}T_B
g T_C) \ . 
\end{equation}
Then we obtain that the
 variation of the
action (\ref{Sgafix})
takes the form (We again presume
that $\chi^\pm,\chi^\xi$ do not 
depend on $J$) 
\begin{equation}
\delta S_{matter}=-\frac{\sqrt{\lambda}}{2\pi}
\int d\sigma^2\sqrt{-\gamma}
\gamma^{\alpha\beta}
\partial_\alpha \epsilon^A
C_{AB}K^{BC}K_{CD}J_\beta^D \ .
\end{equation}
Then the standard arguments
imply an existence of the
current
\begin{equation}\label{ja}
j_{A\beta}=\frac{\sqrt{\lambda}}{2\pi}
 C_{AC}J_\beta^C \  
\end{equation}
that is conserved
\begin{equation}\label{connc}
\partial_\alpha [\sqrt{-\gamma}\gamma^{\alpha
\beta}j_{A\beta}]=0 \  
\end{equation}
on condition that 
all fields obey the equations
of motion. 
Note also that the current  
(\ref{ja})  
$j=j^AT_A$ can be written in the following
form
\begin{equation}\label{Lj}
j=\frac{\sqrt{\lambda}}{2\pi}
dgg^{-1}
=\frac{\sqrt{\lambda}}{2\pi}
gJg^{-1} \ . 
\end{equation}
Then from (\ref{Lj}) we obtain 
important identity
\begin{equation}
dj+j\wedge j=0 
\end{equation}
or explicitly 
\begin{equation}\label{flj}
\partial_\alpha j_\beta^A-
\partial_\beta j_\alpha^A+
j_\alpha^B j_\beta^C
f_{BC}^A=0 \ . 
\end{equation}
The equation (\ref{flj})
 together with (\ref{connc})
allows us to construct non-local
conserved charges using
the iteration procedure presented
 in \cite{Brezin:1979am}. 

To begin with we define the operator
$(D_\alpha)^A_B$ as
\begin{equation}
(D_\alpha)^A_B=
\delta^A_B\partial_\alpha +
j^C_\alpha f_{CB}^A  \ . 
\end{equation}
This operator acts  on an element
$X^A$  as
\begin{equation}
(D_\alpha)^A_BX^B=
\partial_\alpha X^A+
j^C_\alpha f_{CB}^AX^B \ . 
\end{equation}
In the same way we introduce
the operator 
\begin{equation}
(\partial_{\alpha})^A_B=
\delta^A_B\partial_\alpha \ . 
\end{equation}
Then it is easy to see that
\begin{eqnarray}\label{parD}
[(\partial_{\alpha})^A_B(\sqrt{-\gamma}\gamma^{\alpha\beta}
D_{\beta C}^B)-D^A_{\alpha B}(
\sqrt{-\gamma}\gamma^{\alpha\beta}(\partial_{\beta})^B_C)]
X^C=
\partial_\alpha[\sqrt{-\gamma}\gamma^{\alpha\beta}
j_\beta^B]f_{BC}^A X^C=0
\nonumber \\
\end{eqnarray}
due to the fact that $j$ obeys the
equation (\ref{connc}). 

In the same way we can show that
\begin{eqnarray}\label{DD}
[D_{\alpha },D_{\beta }]^A_BX^B
=[\partial_\alpha j^C_\beta-\partial_\beta
j^C_\alpha+j^D_\alpha j^E_\beta f_{DE}^C]
f_{CB}^AX^B=0
\nonumber \\
\end{eqnarray}
due to the equation (\ref{flj}). 
Now we return to the iterative procedure given
in \cite{Brezin:1979am} generalised to the
case of general world-sheet metric. 
Let us consider  conserved
current $j^{(n)}_{\alpha A}$ that
obeys the equation 
\begin{equation}\label{conncn}
\partial_\alpha[\sqrt{-\gamma}
\gamma^{\alpha\beta} j^{(n)}_{\beta A}]=0 \ . 
\end{equation}
Then we can define the function $\xi^{(n)}_A$
in the following way 
\begin{equation}\label{jxi}
j^{(n)}_{\alpha A}=
\gamma_{\alpha\delta}
\epsilon^{\delta\beta}\partial_\beta
\xi^{(n)}_A
\end{equation}
since then (\ref{conncn}) is trivially 
satisfied as follows from 
\begin{equation}
\partial_\alpha[\sqrt{-\gamma}\gamma^{\alpha\beta}
 j^{(n)}_{\beta A}]=
\partial_\alpha[\sqrt{-\gamma} \epsilon^{\alpha\beta}
\partial_\beta \xi^{(n)}_A]
=e^{\alpha\beta}
\partial_\beta\partial_\alpha \xi^{(n)}_A=0 \ , 
\end{equation}
where we have again used the fact that 
\begin{equation}
\epsilon^{\alpha\beta}=\frac{e^{\alpha\beta}}
{\sqrt{-\gamma}} \ , e^{01}=-e^{10}=1 \ . 
\end{equation}
Note also  that the equation (\ref{jxi})
implies
\begin{equation}\label{xij}
\partial_\alpha \xi^{(n)A}=
K^{AB}\epsilon_{\alpha\beta}\gamma^{\beta\gamma}
j_{\gamma B}^{(n)}=
K^{AB}\gamma_{\alpha\beta}\epsilon^{\beta\gamma}
j_{\gamma B}^{(n)}
\end{equation}
using the fact that 
 $\gamma_{\alpha\beta}\epsilon^{\beta\delta}=
 \epsilon_{\alpha\beta}\gamma^{\beta\delta}$.
Then we define current 
\begin{equation}
j_{\alpha A}^{(n+1)}=
K_{AB}(D_{\alpha})^B_C\xi^{C(n)}
=\partial_\alpha \xi^{(n)}_A+
K_{AB}j^{D}_\alpha \xi^{(n)C}
f_{DC}^B \ .
\end{equation}
This current is also conserved
as follows from  
\begin{eqnarray}\label{cocn}
& & \partial_\alpha[
\sqrt{-\gamma}\gamma^{\alpha\beta}
j^{(n+1)}_{\beta A}]=
K_{AB} \partial_\alpha[\sqrt{-\gamma}
\gamma^{\alpha\beta}(D_{\beta})^B_C\xi^{(n)C}]=
K_{AB}(D_\alpha)^B_{C}[\sqrt{-\gamma}\gamma^{\alpha\beta}
\partial_\beta \xi^{(n)C}]=\nonumber \\
&=&
K_{AB}(D_{\alpha})^B_C[\sqrt{-\gamma}\gamma^{\alpha\beta}
K^{CD}\gamma_{\beta\gamma}\epsilon^{\gamma\delta}
j_{\delta D}^{(n)}]
=K_{AB}e^{\beta\gamma}(D_\beta)^B_{C}
(D_{\gamma})^C_D\xi^{(n-1)D}=0 \ , 
\nonumber \\
\end{eqnarray}
where we have used 
(\ref{parD}) and (\ref{DD}). 
The fact that $j^{(n+1)}$ obeys
(\ref{cocn}) implies that 
this iterative procedure generates 
 infinite number of non-local currents
with  initial condition
\begin{equation}
j^{(0)}_{\alpha A}=j_{\alpha A} \ . 
\end{equation}
As an example we determine 
the first order  non-local current.
Firstly from (\ref{xij}) we obtain
\begin{equation}
\xi^{(0)}_A(\sigma,\tau)=
-\int_0^{\sigma}d\sigma' \sqrt{-\gamma}
\gamma^{\tau\delta}j_{\delta A}(\tau,\sigma')+
\int_{-\infty}^\tau
d\tau' \sqrt{-\gamma}\gamma^{\sigma\delta}
j_{\delta A}(\tau',0) \ , 
\end{equation}
where we have used the initial condition
 $\lim_{\tau\rightarrow -\infty} j(\tau)=0$.
Then we obtain
\begin{eqnarray}
j^{(1)}_{\sigma A}(\sigma,\tau)
&=&
\epsilon_{\sigma\alpha}
\gamma^{\alpha\beta}j_{\beta A}(\sigma,\tau)
+K_{AB}j_{\sigma}^D \xi^{(0)C}(\tau,\sigma)f_{DC}^B \ , 
\nonumber \\
j^{(1)}_{\tau A}(\tau,\sigma)&=&
\epsilon_{\tau\alpha}\gamma^{\alpha\beta}
j_{\beta A}(\sigma,\tau)+K_{AB}j_{\tau}^D \xi^{(0)C}(\tau,\sigma)
f_{DC}^B \ . 
\nonumber \\
\end{eqnarray}
In order to define the time-independent
conserved charge we have to carefully specify the
integration  domain due to the finite size
of the string world-sheet. To do this we
use the proposal given recently in 
\cite{Hatsuda:2006ts} and define
the  conserved charge $Q^{(n)}$ as
\begin{equation}\label{defQn}
Q^{(n)}_A=-
\int_{-\infty}^\tau d\tau'\sqrt{-\gamma}\gamma^{\sigma\alpha}
j^{(n)}_{\alpha A}(\tau',0)+
\int_0^{2\pi}d\sigma' \sqrt{-\gamma}
\gamma^{\tau\alpha}j_{\alpha A}^{(n)}(\sigma',\tau)+
\int_{-\infty}^\tau d\tau' \sqrt{-\gamma}
\gamma^{\sigma\alpha}j^{(n)}_{\alpha A}(\tau',2\pi) \ . 
\end{equation}
Then using the conservation of 
$j_\alpha^{(n)}$ and integration 
by parts we can show that the 
charge $Q^{(n)}$ defined
in  (\ref{defQn}) is conserved
\begin{eqnarray}
\frac{dQ^{(n)}_A}{d\tau}=-
\sqrt{-\gamma}\gamma^{\sigma\alpha}
j^{(n)}_{\alpha A}(\tau,0)
-\int_0^{2\pi}d\sigma'
\partial_{\sigma'}[\sqrt{-\gamma}
\gamma^{\sigma\alpha}j_{\alpha A}^{(n)}(\sigma',\tau)]
+\sqrt{-\gamma}\gamma^{\sigma\alpha}
j^{(n)}_{\alpha A}(\tau,2\pi)=0 \ . 
\nonumber \\
\end{eqnarray}
We mean that it is very satisfactory
result  that
we can define non-local conserved charges
in case of finite domain as well. This
fact could be useful for further study of the
integrability of the string theory on
$AdS_5\times S^5$.
\subsection{BRST invariance of the
conserved non-local charges}
Now we are going to show that the
non-local charges given above are
BRST invariant as well. To do this
we have to determine the BRST variation
of $\xi^{(n)}_A$. It is easy to see
that this is equal to
\begin{equation}
\delta_{Q}\xi^{(n)}_A=
C^\beta \partial_\beta \xi^{(n)}_A
\end{equation}
since then
\begin{eqnarray}
\delta_{Q}(j^{(n)}_{\alpha A})&=&
\delta_Q(\gamma_{\alpha\delta}
\epsilon^{\delta\beta})\partial_\beta
\xi^{(n)}_A+
\gamma_{\alpha\delta}
\epsilon^{\delta\beta}\partial_\beta
\delta_{Q}(\xi^{(n)}_A)=\nonumber \\
&=&C^\beta \partial_\beta j^{(n)}_\alpha+
j_{\beta}^{(n)}\partial_\alpha C^\beta 
\nonumber \\
\end{eqnarray} 
that is a correct form of the BRST transformation
of the current $j_\alpha$ as follows
from (\ref{deltaG}). 
Then the BRST variation of the current
$j^{(n+1)}_{A\alpha }$ is equal to
\begin{eqnarray}
\delta_{Q}j^{(n+1)}_{A\alpha}&=&
\partial_\alpha \delta_{Q}(\xi^{(n)}_A)+
K_{AB}\delta_{Q}(j^D_\alpha)\xi^{(n)C}
f_{DC}^B+K_{AB}
j^{D}_\alpha \delta_{Q}(\xi^{(n)C})
f_{DC}^B=\nonumber \\
&=&\partial_\alpha C^\gamma
(\partial_\gamma \xi^{(n)}_{A}+
K_{AB}j_{\gamma}^{D}\xi^{(n)C}f_{DC}^B)+
C^\gamma\partial_\gamma (\partial_\alpha
\xi_{A}^{(n)}+
K_{AB}j_\alpha^D \xi^{(n)C}f_{DC}^B)=
\nonumber \\
&=&\partial_\alpha C^\beta j^{(n+1)}_{A\beta}+
\partial_\beta j_{A\alpha}^{(n+1)} C^\beta \ . 
\nonumber \\
\end{eqnarray}
Then using (\ref{heBRST}) we easily 
get
\begin{eqnarray}\label{heBRSTn}
\delta_{Q}(\sqrt{-\gamma}\gamma^{\tau\alpha}
j_{A\alpha}^{(n)})&=&
\partial_\tau (\sqrt{-\gamma}j^{(n)\tau}_A)C^\tau+
\partial_\sigma (\sqrt{-\gamma}j^{(n)\tau}_AC^\sigma)
-\sqrt{-\gamma} j^{(n)\sigma}_A \partial_\sigma C^\tau \ , 
\nonumber \\
\delta_{Q}(\sqrt{-\gamma}
\gamma^{\sigma\alpha}j_{A\alpha}^{(n)})&=&
\partial_\sigma(\sqrt{-\gamma} j^{(n)\sigma}_A)C^\sigma+
\partial_\tau (\sqrt{-\gamma} j^{(n)\sigma}_A C^\tau)-
\sqrt{-\gamma}j^{(n)\tau}_A\partial_\tau C^\sigma \ . 
\nonumber \\
\end{eqnarray}
Let us now calculate following
Poisson bracket
\begin{eqnarray}\label{Qnl}
\pb{Q^{(n)}_A,Q}=
-\int_{-\infty}^\tau d\tau'\delta_{Q}(
\sqrt{-\gamma}\gamma^{\sigma\alpha}
j^{(n)}_{\alpha A}(\tau',0))+\nonumber \\
\int_0^{2\pi}d\sigma' \delta_{Q}( \sqrt{-\gamma}
\gamma^{\tau\alpha}j_{\alpha A}^{(n)}(\sigma',\tau))+
\int_{-\infty}^\tau d\tau'\delta_{Q}( \sqrt{-\gamma}
\gamma^{\sigma\alpha}j^{(n)}_{\alpha A}(\tau',2\pi)) \ . 
\end{eqnarray}
Using (\ref{heBRSTn}) the
first expression above can
be calculated as 
\begin{eqnarray}\label{nl1}
& &-\int_{-\infty}^\tau d\tau'\delta_{Q}(
\sqrt{-\gamma}\gamma^{\sigma\alpha}
j^{(n)}_{\alpha A}(\tau',0))=
\nonumber \\
&=&-(\sqrt{-\gamma} j^{(n)\sigma}_A C^\tau)(\tau,0)+
(\sqrt{-\gamma} j^{(n)\tau}_AC^\sigma)(\tau,0) \ . 
\nonumber \\
\end{eqnarray}
In the same way we proceed with the
third term in (\ref{Qnl}) and we
obtain the result
\begin{eqnarray}\label{nl2}
& &\int_{-\infty}^\tau d\tau'\delta_{Q}(
\sqrt{-\gamma}\gamma^{\sigma\alpha}
j^{(n)}_{\alpha A}(\tau',2\pi))=
\nonumber \\
&=&(\sqrt{-\gamma} j^{(n)\sigma}_A C^\tau)(\tau,2\pi)-
(\sqrt{-\gamma} j^{(n)\tau}_AC^\sigma)(\tau,2\pi) \ . 
\nonumber \\ 
\end{eqnarray}
Finally we will calculate
the second expression in 
(\ref{Qnl}) 
\begin{eqnarray}\label{nl3}
& &\int_0^{2\pi}d\sigma' \delta_{Q}( \sqrt{-\gamma}
\gamma^{\tau\alpha}j_{\alpha A}^{(n)}(\sigma',\tau))=
\nonumber \\
&=&\int_0^{2\pi}d\sigma'
(\partial_\tau (\sqrt{-\gamma}j^{(n)\tau}_A)C^\tau+
\partial_{\sigma'} (\sqrt{-\gamma}j^{(n)\tau}C^\sigma)
-\sqrt{-\gamma} j^{(n)\sigma} \partial_{\sigma'} C^\tau)=
\nonumber \\
&=&-\sqrt{-\gamma}j^{(n)\sigma}_AC^\tau(\tau,2\pi)+
\sqrt{-\gamma}j^{(n)\sigma}_AC^\tau(\tau,0)+
\nonumber \\
&+& 
\sqrt{-\gamma}j^{(n)\tau}_AC^\sigma(\tau,2\pi)-
\sqrt{-\gamma}j^{(n)\tau}_AC^\sigma(\tau,0) \ . 
\nonumber \\
\end{eqnarray}
Then collecting (\ref{nl1}),(\ref{nl2})
and (\ref{nl3}) we obtain the final
result 
\begin{equation}
\pb{Q^{(n)}_A,Q}=0 \ . 
\end{equation}
In other words the non-local conserved
charges are BRST invariant. 
\section{BRST invariance of the
monodromy matrix}\label{fifth}
In this section we  show that
 monodromy matrix of the principal
model is  BRST invariant as well. 
We mean that this result  could be useful
for further study of the properties
of the  bosonic string on 
$AdS_5\times S^5$. 
For example, it is well know that the
trace of the monodromy matrix is a generator
of the local conserved charges. Then
the BRST invariance of the monodromy
matrix immediately implies 
BRST invariance of these
charges as well. Moreover, it is well
know that the
 monodromy matrix  is the generator
of non-local charges and hence
these charges are BRST invariant. 
We mean that the fact 
that local and non-local charges
are BRST invariant  support their
physical importance for the description
of the state of the bosonic string
on $AdS_5\times S^5$. 

Before we proceed to the study of the
BRST properties of the monodromy matrix
we review its properties with respect
to phase space and canonical Poisson
brackets.  
\subsection{Hamiltonian analysis
of the monodromy matrix}
In this section we review the main
properties of the monodromy matrix.
We closely follow 
\cite{Faddeev:1987ph,deVega:1983gy,Izergin:1980pe}.
We also construct,
following  \cite{Hatsuda:2006ts},
 the monodromy matrix
on the string 
 world-sheet  that has finite spatial
extend.

The monodromy matrix is defined
as 
\begin{equation}\label{Tdef}
\mT(\gamma,\Lambda)=
P\exp \left(\int_\gamma 
\hJ_\alpha(\Lambda) \frac{dx^\alpha}{d\xi}\right) \ , 
\end{equation}
where $\gamma=(x^\alpha(\xi)\equiv (\tau'(\xi),
\sigma'(\xi))$ is a curve on
 two dimensional cylinder, $\hJ_\alpha \frac{dx^\alpha}{d\xi}$
is an embedding of the Lax connection with 
spectral parameter $\Lambda$ on the curve
$\gamma$  and $\xi$ is a
parameter that labels points on given curve. Finally
 $P$ in
(\ref{Tdef}) means  path ordering. 

In order to define the time-independent monodromy
matrix on string world-sheet we consider
following form of the curve $\gamma$
\begin{eqnarray}\label{gammaf}
\gamma&=&\gamma_1(-\infty,0,\tau,0)+
\gamma_2(\tau,0,\tau,2\pi)+
\gamma_3(\tau,2\pi,-\infty,2\pi) \ ,  \nonumber \\
\gamma_1&=&(\tau'=\xi,\sigma'=0,\xi\in (-\infty,\tau)) \ , 
\nonumber \\
\gamma_2&=&(\tau'=\tau,\sigma'=\xi, \xi\in
(0,2\pi)) \ , \nonumber \\
\gamma_3&=&(\tau'=\xi, \sigma'=2\pi,\xi\in (\tau,-\infty)) 
\ . \nonumber \\
\end{eqnarray}
Then for the   curve (\ref{gammaf})
the monodromy matrix   $\mT$ takes the form
\begin{equation}\label{mTed}
\mT(\Lambda)=
P\exp(\int_{-\infty}^\tau 
d\tau' \hJ_\tau(\tau',0)
+\int_0^{2\pi}d\sigma'
\hJ_\sigma (\tau,\sigma')
+\int_\tau^{-\infty}
d\tau'\hJ_\tau(\tau',2\pi)) \ . 
\end{equation}
 Let us now 
 calculate the Poisson bracket
between any local quantity (for
example $H$ or $Q$) and the monodromy
matrix for general form of the
curve  $\gamma$. Using
the definition of Poisson bracket
we obtain
\begin{eqnarray}\label{pbXTg}
\pb{X,\mT_{ij}(\gamma,\Lambda)}=
\int_0^{2\pi}
d\sigma \left(\frac{\delta X}{\delta
J^A_\sigma(\sigma)}
\frac{\delta \mT_{ij}(\gamma,\Lambda)}
{\delta \Pi_A(\sigma)}-
\frac{\delta X}{\delta
\Pi_A(\sigma)}\frac{\delta \mT_{ij}(\gamma,\Lambda)}
{\delta J^A_\sigma(\sigma)}\right) \ , 
\nonumber \\
\end{eqnarray}
where $(ij)$ label the matrix indices of $\mT$.
To proceed we use the fact that 
(For simplicity we omit the symbol $\Lambda$)
\begin{eqnarray}
\frac{\delta \mT_{ij}(\gamma)}
{\delta J^B_\sigma(\sigma)}&=&
\int_\gamma d\xi \frac{\delta 
\hJ^A_\alpha(x(\xi))}{\delta 
J^B_\sigma(\sigma)}
\frac{dx^\alpha}{d\xi}\frac{\delta \mT_{ij}(\gamma)}
{\delta \hJ^A(x(\xi))}
 \ , \nonumber \\
\frac{\delta \mT_{ij}(\gamma)}
{\delta \Pi_B(\sigma)}&=&
\int_\gamma d\xi \frac{\delta 
\hJ^A_\alpha(x(\xi))}{\delta 
\Pi_B(\sigma)}\frac{dx^\alpha}{d\xi}
\frac{\delta \mT_{ij}
(\gamma)}{\delta \hJ^A(x(\xi))} 
 \ . \nonumber \\
\end{eqnarray}
Then we can rewrite
(\ref{pbXTg}) into the form
\begin{eqnarray}\label{pbXT}
\pb{X,T_{ij}(\gamma)}&=&
\int_\gamma d\xi
\frac{\delta \mT_{ij}(\gamma)}
{\delta \hJ^A(x(\xi))}\frac{dx^\alpha}{d\xi}
\int_0^{2\pi}
d\sigma \left(\frac{\delta X}{\delta J^B_\sigma(\sigma)}
\frac{\delta \hJ^A_\alpha (x(\xi))}
{\delta \Pi_B(\sigma)}-
\frac{\delta X}{\delta \Pi_B(\sigma)}
\frac{\delta \hJ^A_\alpha (x(\xi))}
{\delta J^B_\sigma(\sigma)}\right)=
\nonumber \\
&=&
\int_\gamma d\xi
\frac{\delta \mT_{ij}(\gamma)}
{\delta \hJ^A(x(\xi))}\frac{dx^\alpha}{d\xi}
\pb{X,\hJ^A_\alpha(x(\xi))} \ . \nonumber \\
\end{eqnarray}
As the next step we determine
the variation of $\mT_{ij}$ with respect
to $\hJ$. If we generalize
the arguments given in 
\cite{deVega:1983gy,Izergin:1980pe}
to the case of arbitrary curve $\gamma$
we obtain 
\begin{eqnarray}\label{deltamT}
\delta \mT(\gamma)=
\int_\gamma d\xi
\mT(\gamma'<\gamma(\xi))
\delta\hJ_\alpha(\xi)\frac{dx^\alpha}{d\xi}
\mT(\gamma'>\gamma(\xi))  \ , \nonumber \\
\end{eqnarray}
where we have introduced following
notation. $\gamma'>\gamma(\xi)$ means
the part of the  curve $\gamma$ 
that starts at the point $\gamma(\xi)$
and ends at the final point of
curve $\gamma(\xi_f)$. In the same way 
$\gamma'<\gamma(\xi)$ denotes the
curve that starts at the initial
point of $\gamma(\xi_i)$ and ends at
the point
$\gamma(\xi)$. Finally 
 $\mT(\gamma'>\gamma(\xi))\ , 
\mT(\gamma'<\gamma(\xi))$ denote 
 monodromy matrices evaluated on 
the curves $\gamma'>\gamma(\xi)$,
$\gamma'<\gamma(\xi)$ respectively. 
Then if we use the fact that
 $\hJ=\hJ^A T_A$ we
can rewrite  (\ref{deltamT}) as
\begin{eqnarray}
\delta \mT_{ij}(\gamma)&=&
\int_\gamma d\xi \mT(\gamma'<\gamma(\xi))_{ik}
\delta (\hJ_{\alpha})_{kl}\frac{dx^\alpha}
{d\xi}\mT_{lj}(\gamma'>\gamma(\xi))=
\nonumber \\
&=&\int_\gamma d\xi
\delta \hJ^A_\alpha(\gamma(\xi))\frac{dx^\alpha}{d\xi}
\mT_{ik}(\gamma'<\gamma(\xi))_{kl}(T_A)_{kl}
\mT_{lj}(\gamma'>\gamma(\xi))
\nonumber \\
\end{eqnarray}
and finally
\begin{equation}\label{deltaTf}
\frac{\delta \mT_{ij}(\gamma)}
{\delta \hJ^A(\gamma(\xi))}=
\mT_{ik}(\gamma'<\gamma(\xi))
(T_A)_{kl}\mT_{lj}(\gamma'>\gamma(\xi)) \ ,  
\nonumber \\ 
\end{equation}
where $\hJ^A(\xi)=\hJ^A_\alpha(x(\xi))
\frac{dx^\alpha}{d\xi}$.
Then using (\ref{deltaTf}) the
Poisson bracket (\ref{pbXT}) takes
the final form
\begin{equation}
\pb{X,\mT_{ij}(\gamma)}
=\int_\gamma d\xi
\mT_{ik}(\gamma'<\gamma(\xi))
(T_A)_{kl}
\mT_{lj}(\gamma'>\gamma(\xi))
\frac{dx^\alpha}{d\xi}
\pb{X,\hJ^A_\alpha(\gamma(\xi))}
\end{equation}
or in matrix notation
\begin{equation}
\pb{X,\mT(\gamma)}\label{pbXTfm}
=\int_\gamma d\xi
\mT(\gamma'<\gamma(\xi))
\pb{X,\hJ_\alpha(\gamma(\xi))}\frac{dx^\alpha}{d\xi}
\mT(\gamma'>\gamma(\xi)) \ . 
\end{equation}
With the help of this formula we
show that $\mT$ evaluated on the
curve (\ref{gammaf}) commutes with
Hamiltonian on condition that
$\hJ$ is Lax pair. To do this
we use the fact that
the  time evolution of any component
of the Lax connection is governed
by the equation 
\begin{equation}
\partial_\tau\hJ_\alpha(\sigma,\tau)=
-\pb{H,\hJ_\alpha(\sigma,\tau)} \ . 
\end{equation}
Then the equation (\ref{pbXTfm}) for
$X=H$ implies
\begin{eqnarray}\label{mTH}
-\pb{\mT,H}=
\int_{\gamma}
d\xi 
\mT(\gamma'<\gamma(\xi))
\partial_\tau\hJ_\alpha(\gamma(\xi))
\frac{dx^\alpha}{d\xi}
\mT(\gamma'>\gamma(\xi))=\nonumber \\
=\int_{-\infty}^\tau
d\tau'\mT(-\infty,0,\tau',0)
\partial_{\tau'}\hJ_\tau(\tau',0)
\mT(\tau',0,-\infty,2\pi)+
\nonumber \\
+\int_0^{2\pi}d\sigma\mT(-\infty,0,\tau,\sigma)
\partial_\tau \hJ_\sigma(\tau,
\sigma)\mT(\tau,\sigma,-\infty,2\pi)+
\nonumber \\
+\int_\tau^{-\infty}d\tau'
\mT(-\infty,0,\tau',2\pi)
\partial_{\tau'}
 \hJ_\tau(\tau',2\pi)
\mT(\tau,2\pi,-\infty,2\pi) \ . 
\nonumber \\
\end{eqnarray}
The first term in (\ref{mTH}) can
evaluated using the  integration by
parts and we obtain
\begin{eqnarray}\label{ev1p}
& &\int_{-\infty}^\tau
d\tau'\mT(-\infty,0,\tau',0)
\partial_{\tau'}\hJ_\tau(\tau',0)
\mT(\tau',0,-\infty,2\pi)=\nonumber \\
&=&
\mT(-\infty,0,\tau,0)
\hJ_\tau(\tau,0)
\mT(\tau,0,-\infty,2\pi) \ , \nonumber \\
\end{eqnarray}
where we have  used the
fact that
\begin{eqnarray}\label{mTXY}
\frac{\partial\mT(\gamma)}
{\partial{Y^\alpha}}&=&\mT(\gamma)\hJ_\alpha(Y) \ ,
\quad Y\equiv \gamma(\xi_f) \ ,  
\nonumber \\
\frac{\partial\mT(\gamma)}
{\partial X^\alpha}&=&-\hJ_\alpha(X)
\mT(\gamma) \ , 
\quad X\equiv\gamma(\xi_i) \ , 
\nonumber \\
\end{eqnarray}
where $X^\alpha$ is the initial point
of the curve $\gamma$ and
$Y^\alpha$ is the final point of the curve. 
In the same way we can show that
\begin{eqnarray}\label{ev3p}
& &\int_\tau^{-\infty}d\tau'
\mT(-\infty,0,\tau',2\pi)
\partial_{\tau'}
 \hJ_\tau(\tau',2\pi)
\mT(\tau,2\pi,-\infty,2\pi)
=\nonumber \\
&=&-\mT(-\infty,0,\tau,2\pi)
 \hJ_\tau(\tau,2\pi)
\mT(\tau,2\pi,-\infty,2\pi) \ . 
\nonumber \\
\end{eqnarray}
Finally we evaluate 
the second expression in 
(\ref{mTH}) using the fact that
$\hJ$ is Lax connection. Explicitly
we get
\begin{eqnarray}\label{ev2p}
& &\int_0^{2\pi}d\sigma
\mT(-\infty,0,\tau,\sigma)
\partial_\tau \hJ_\sigma(\tau,
\sigma)\mT(\tau,\sigma,-\infty,2\pi)=
\nonumber \\
&=&\int_0^{2\pi}d\sigma\mT(-\infty,0,\tau,\sigma)
(\partial_{\sigma} \hJ_\tau(\tau,\sigma)-
[\hJ_\tau,\hJ_\sigma](\tau,\sigma))
\mT(\tau,\sigma,-\infty,2\pi)
\nonumber \\
&=&\mT(-\infty,0,\tau,2\pi)
 \hJ_\tau(\tau,2\pi)\mT(\tau,2\pi,-\infty,2\pi)
-\mT(-\infty,0,\tau,0)
 \hJ_\tau(\tau,0)\mT(\tau,0,-\infty,2\pi)
\ , \nonumber \\
\end{eqnarray}
where we have performed
integration by parts and  used
(\ref{mTXY}). If we now combine
(\ref{ev2p}) together
with (\ref{ev1p}) and (\ref{ev3p})
we obtain that
\begin{equation}
\pb{\mT(\Lambda),H}=0 \ . 
\end{equation}
In other words the monodromy matrix
defined in (\ref{mTed}) is time
independent and hence it can 
be considered as generator
of infinite number of local and
non-local charges.

Finally we show that the monodromy
matrix (\ref{mTed}) is BRST invariant. 
Recall that the  BRST
transformation of the Lax connection
takes the form  
\begin{equation}\label{pbhJQ}
\delta_Q\hJ_\alpha=
\pb{\hJ_\alpha,Q}
=\partial_\alpha C^\beta \hJ_\beta+
\partial_\alpha \hJ_\beta C^\beta \ .
\end{equation}
Now using (\ref{pbXTfm}) we obtain
\begin{eqnarray}\label{pbmTQ}
\pb{\mT(\gamma),Q}
&=&\int_\gamma d\xi
\mT(\gamma'<\gamma(\xi))
\pb{\hJ_\alpha(x(\xi)),Q}\frac{dx^\alpha}{d\xi}
\mT(\gamma'>\gamma(\xi))=
\nonumber \\
&=&\int_{-\infty}^\tau
d\tau'\mT(-\infty,0,\tau',0)
\pb{\hJ_\tau(\tau',0),Q}
\mT(\tau',0,-\infty,2\pi)+
\nonumber \\
&+&\int_0^{2\pi}d\sigma\mT(-\infty,0,\tau,\sigma)
\pb{\hJ_\sigma(\tau,
\sigma),Q}\mT(\tau,\sigma,-\infty,2\pi)+
\nonumber \\
&+&\int_\tau^{-\infty}d\tau'
\mT(-\infty,0,\tau',2\pi)
\pb{ \hJ_\tau(\tau',2\pi),Q}
\mT(\tau,2\pi,-\infty,2\pi) \ . 
\nonumber \\
\end{eqnarray}
Using (\ref{pbhJQ}) we can
calculate the first term 
in (\ref{pbmTQ}) as
\begin{eqnarray}\label{pbmTQ1}
& &\int_{-\infty}^\tau
d\tau'\mT(-\infty,0,\tau',0)
\pb{\hJ_\tau(\tau',0),Q}
\mT(\tau',0,-\infty,2\pi)= \nonumber \\
&=&\mT(-\infty,0,\tau,0)C^\alpha \hJ_\alpha
\mT(\tau,0,-\infty,0) \ , \nonumber \\
\end{eqnarray}
where we have performed the integration 
by parts and used  
(\ref{mTXY}) and also the fact that
Lax connection is flat. In 
the same way we obtain
\begin{eqnarray}\label{pbmTQ2}
& &\int_\tau^{-\infty}d\tau'
\mT(-\infty,0,\tau',2\pi)
\pb{ \hJ_\tau(\tau',2\pi),Q}
\mT(\tau,2\pi,-\infty,2\pi)=
\nonumber \\
&=&-\mT(-\infty,2\pi,\tau,2\pi)C^\alpha \hJ_\alpha
\mT(\tau,2\pi,-\infty,2\pi) \ . \nonumber \\
\end{eqnarray}
Finally we will calculate second
term in (\ref{pbmTQ}) and we get 
\begin{eqnarray}\label{pbmTQ3}
& &\int_0^{2\pi}d\sigma\mT(-\infty,0,\tau,\sigma)
\pb{\hJ_\sigma(\tau,
\sigma),Q}\mT(\tau,\sigma,-\infty,2\pi)=
\nonumber \\
&=&\mT(-\infty,0,\tau,2\pi)C^\alpha
\hJ_\alpha(\tau,2\pi)\mT(\tau,\sigma,
\infty,2\pi)-
\nonumber \\
&-&
\mT(-\infty,0,\tau,0)C^\alpha
\hJ_\alpha(\tau,0)\mT(\tau,0,
\infty,2\pi) \ . \nonumber \\
\end{eqnarray}
Collecting (\ref{pbmTQ1}),
(\ref{pbmTQ2}) and 
(\ref{pbmTQ3}) together we
obtain
\begin{equation}
\pb{\mT,Q}=0 \ . 
\end{equation} 
In other words, the monodromy
matrix is BRST invariant.

\section*{Acknowledgements}

I would like to thank 
Institute of Theoretical Physics and Astrophysics,
Masaryk University, Brno   for hospitality
where part  of  this work has been done. 
This work  was supported in part by the Czech Ministry of
Education under Contract No. MSM
0021622409, by INFN, by the MIUR-COFIN
contract 2003-023852 and , by the EU
contracts MRTN-CT-2004-503369 and
MRTN-CT-2004-512194, by the INTAS
contract 03-516346 and by the NATO
grant PST.CLG.978785.

\end{document}